\documentclass[11pt,a4paper]{article}
 \pdfoutput=1
\usepackage{jcappub}
\usepackage{amsmath}
\usepackage{amsfonts,color}
\usepackage{amssymb,float}
\usepackage{mathtools}
\usepackage[utf8]{inputenc}
\usepackage{url}
\usepackage{graphicx}
\usepackage{refstyle}
\usepackage{wasysym}
\usepackage{tabularx}
\usepackage{accents}
\usepackage{graphicx}
\usepackage{color}
\usepackage[dvipsnames]{xcolor}
\usepackage{hyperref}

\hypersetup{
    colorlinks=true,
    linkcolor=blue,
    filecolor=magenta,      
    citecolor=red
}

\usepackage{subfigure}		
\usepackage{hyperref} 

\makeatletter
\long\def\dddddot#1{%
  {\mathop {#1}\limits ^{\vbox to-1.4\ex@ {\kern -\tw@ \ex@ \hbox {\normalfont .....}\vss }}}%
}
\long\def\multidots#1#2{%
  \count@=0
  {{\mathop {#2}\limits ^{\vbox to-1.4\ex@ {\kern -\tw@ \ex@ \hbox {\normalfont %
  \loop%
  \ifnum#1>\count@%
  .%
  \advance\count@ by1%
  \repeat%
  }\vss }}}}%
}
\makeatother

\title{\boldmath 
Towards a model-independent reconstruction approach for late-time Hubble data
}

\author[a]{Reginald Christian Bernardo,}
\author[b,c]{Jackson Levi Said}

\affiliation[a]{National Institute of Physics, University of the Philippines Diliman, Quezon City 1101, Philippines}
\affiliation[b]{Institute of Space Sciences and Astronomy, University of Malta, Malta, MSD 2080}
\affiliation[c]{Department of Physics, University of Malta, Malta, MSD 2080}

\emailAdd{rbernardo@nip.upd.edu.ph}
\emailAdd{jackson.said@um.edu.mt}

\abstract{
\textbf{Gaussian processes offers a convenient way to perform nonparametric reconstructions of observational data assuming only a kernel which describes the covariance between neighbouring points in a data set. We approach the ambiguity in the choice of kernel in Gaussian processes with two methods -- (a) approximate Bayesian computation with sequential Monte Carlo sampling and (b) genetic algorithm -- and use the overall resulting method to reconstruct the cosmic chronometers and supernovae type Ia data sets.
The results have shown that the Mat\'{e}rn$\left( \nu = 5/2 \right)$ kernel emerges on top of the two-hyperparameter family of kernels for both cosmological data sets. On the other hand, we use the genetic algorithm in order to select a most naturally-fit kernel among a competitive pool made up of a ten-hyperparameters class of kernels. Imposing a Bayesian information criterion-inspired measure of the fitness, the results have shown that a hybrid of the Radial Basis Function and the Mat\'{e}rn$\left( \nu = 5/2 \right)$ kernel best represented both data sets. The kernel selection problem is not totally closed and may benefit from further analysis using other strategies to resolve an optimal kernel for a particular data set.}
}

\begin{document}

\maketitle
\flushbottom

\section{Introduction}
\label{sec:intro}

The value of the Hubble parameter at current times remains a perplexing issue that has led to a growing tension in predicted values of $H_0$ \cite{DiValentino:2020zio,Riess:2020sih,Pesce:2020xfe,deJaeger:2020zpb}. The tension is borne out of the confrontation of measurements from the early Universe with that of the late-time Universe \cite{Bernal:2016gxb}. Early Universe measurements principally rely on a cosmological model in order to make measurements which is regularly assumed to be vanilla $\Lambda$CDM cosmology \cite{Peebles:2002gy,Copeland:2006wr}, and produces generically lower values of $H_0$ \cite{Aghanim:2018eyx,Ade:2015xua}. On the other hand, late-time observations do not rely on a fiducial cosmological model and can make accurate predictions of the value of $H_0$ without assuming a particular cosmology \cite{Riess:2019cxk,Wong:2019kwg}. The Hubble tension has prompted a mammoth effort in the search for possible resolutions to this problem, which range from issue in systematic errors in the various calculations of $H_0$ to the reconsidering of possible modification beyond $\Lambda$CDM \cite{DiValentino:2021izs,DiValentino:2020zio, Handley:2019tkm}. $\Lambda$CDM can be modified by either extending its matter content beyond the standard model or by changing the underlying gravitational theory \cite{Bull:2015stt,Clifton:2011jh,Capozziello:2011et,Saridakis:2021lqd}. These efforts can reduce the Hubble tension but have a tendency to occasionally create other problems.

In this background, we explore the increasingly popular Gaussian processes (GP) regression method \cite{rasmussen:2003} which is used to reconstruct observational data in a nonparametric way assuming that each point in the data set is part of a stochastic process. Each stochastic process is then related together using a covariance function (or kernel) through which an entire range can be simulated. This approach has been extensively used in a number of setting related to expansion data \cite{Gomez-Valent:2018hwc,Colgain:2021ngq,Yennapureddy:2017vvb,Li:2019nux,Seikel2012,Seikel:2013fda}, $f\sigma_8$ data \cite{Benisty:2020kdt}, and gravitational wave analysis \cite{Belgacem:2019zzu,Moore:2015sza,Canas-Herrera:2021qxs}. In addition to this, GP has been used to approach the inverse problem of extensions to gravity, namely, to use observational data to determine a general form of a Lagrangian function. In Refs.~\cite{Briffa:2020qli,Cai:2019bdh} this was used to determine the background evolution of $f(T)$ gravity, which was then extended to incorporate growth data in Ref.~\cite{LeviSaid:2021yat}. While in Refs.~\cite{Reyes:2021owe,Bernardo:2021qhu} this approach was used for regular Horndeski gravity wherein the impact of the scalar field was studied. Finally, in Ref.~\cite{Cai:2015zoa} the possibility of interacting dark energy and dark matter was explored in further detail.

GP has the potential to produce reconstructions of data sets across wide stretches of regions. However, GP suffers from a major deficiency in the selection of kernel under which the GP regression takes place. This has led to a kernel tension since different kernels lead to different reconstructions. While GP is model-independent in the cosmological context, it does rely on a some statistical background in order to determine an appropriate kernel choice. In the Hubble diagram context, this can lead to slightly different values for the $H_0$ parameter as shown in previous studies as well \cite{Gomez-Valent:2018hwc, Haridasu:2018gqm, Yu:2017iju, Escamilla-Rivera:2021rbe}. To visualize this issue better, consider the following kernels (more details in Sec.~\ref{sec:gaussian_processes}): Radial basis function (RBF), Rational quadratic (RQ), Mat\'{e}rn with index $\nu = x/2$ (M$x$2). In Table~\ref{tab:H0_kernel_tension}, we illustrate this kernel tension for cosmic chronometer data (more details in Sec.~\ref{sec:gaussian_processes}) in terms of the predicted value of $H_0$ for these kernels. It is important to note that the values of $H_0$ predicted using the various kernels in the GP analyses in Table~\ref{tab:H0_kernel_tension} are all within 1$\sigma$ of each other making the statistical tension for $H_0$ very mild. However, it would be interesting to consider approaches to differentiating between these values and how they approach the observational data.

\begin{table}[h!]
\center
\caption{$H_0$ predictions obtained with optimization-based GP reconstruction of the bare cosmic chronometers data set for the commonly-used kernels. RBF stands for the radial basis function (Eq. (\ref{eq:rbf})), RQ for the rational quadratic (Eq. (\ref{eq:rq})), and M$x$2 for the Mat\'{e}rn ($\nu = x/2$) kernel (Eq. (\ref{eq:matern})).}
\begin{tabular}{| c | c |}
\hline
$\phantom{ \dfrac{1}{1} }$ Kernel $\phantom{ \dfrac{1}{1} }$ & $H_0$ $\left[\text{km} \ \text{s}^{-1} \text{Mpc}^{-1} \right]$ \\ \hline\hline
RBF & $67.4 \pm 4.7$ \\ \hline
RQ  & $71.0 \pm 5.6$ \\ \hline
M52  & $68.9 \pm 5.4$ \\ \hline
M72 & $68.7 \pm 5.2$ \\ \hline
\end{tabular}
\label{tab:H0_kernel_tension}
\end{table}

There have been a number of interesting works related to exploring this issue in the literature. Ref.~\cite{Gomez-Valent:2018hwc} offers a comprehensive study in how GP is used to reconstruct expansion data as well as introducing important ways to account for error propagation. In Ref.~\cite{Escamilla-Rivera:2021rbe} GP was contrasted with \textbf{Lo}cally w\textbf{e}ighted \textbf{S}catterplot \textbf{S}moothing together with \textbf{Sim}ulation and \textbf{ex}trapolation method (LOESS-Simex) using a number of approaches to characterizing the performance of each reconstruction approach. Another interesting approach to the kernel tension problem is Ref.~\cite{Haridasu:2018gqm} where a multi-task Gaussian process approach is taken wherein kernels can be used in a joint fashion to conduct a combined learning approach to reconstructing observational data. This is very promising and may further reduce the uncertainties associated with GP regression. On the other hand, Ref.~\cite{pmlr-v64-malkomes_bayesian_2016} proposes to directly compare kernels against the original data and to select the best performing kernels by taking a Bayesian path to optimization. 

In this work, we explore a novel approach to solving the kernel tension problem in the late-time Universe using the approximate Bayesian computation (ABC) based on the sequential Monte Carlo (SMC) algorithm. In ABC, the calculation of the likelihood in Markov Chain Monte Carlo (MCMC) is replaced by a comparison calculation between simulated and original data, this then quantifies the strength of an iteration. Thus, given data $D_0$ and a posterior distribution $P(\theta)$ for a parameter $\theta$, the goal of ABC is to approximate the posterior distribution given $D_0$, namely $P(\theta\vert D_0) \propto f(D_0\vert \theta)P(\theta)$, where $f(D_0\vert \theta)$ is the likelihood of $\theta$ for data $D_0$ \cite{10.3389/fbuil.2017.00052}. Combined with a sequential Monte Carlo sampler means that ABC-SMC can be competitive with MCMC while evading the common intractable problem of determining likelihood functions \cite{2009arXiv0911.1705T}. The ABC-SMC method has been successfully applied to various problems in astrophysics and cosmology \cite{Akeret:2015uha, 2013ApJ...764..116W, 2017A&C....19...16J, Ishida:2015wla}. 

ABC-SMC offers a very interesting approach to determining the best GP kernel given a set of possible choices. However, we are also interested in confronting the broader problem of determining a kernel that best approximates observational data given a base set of kernels which can be combined. In model space, the possible kernels that may be selected to reconstruct data now becomes infinite making the problem possibly almost impossible to resolve. To make the problem tractable, we employ the use of genetic algorithms (GA) to determine possible kernel solutions that best approximate observational data \cite{10.5555/534133}. GAs are extremely useful in solving problems of this nature. They are loosely modelled on the principles of evolution via natural selection along with generational mutations. Thus, we use GAs to eliminate kernel constructions that do not meet certain criteria while allowing high `fitness' kernels to survive and mutate. GAs have shown promise \cite{2012arXiv1202.1643R} in a number of areas in cosmology, for instance in Ref.~\cite{Bogdanos:2009ib} GAs are used to determine best-fit nonparametric models of the dark energy equation of state which is then extended to other expansion parameters in Ref.~\cite{Arjona:2019fwb}. Finally, GAs are also used to perform analyses on growth data in Ref.~\cite{2012JCAP...11..033N} where the evolution of $f\sigma_8$ was studied.

In the present work, we first review GP regression in Sec.~\ref{sec:gaussian_processes} where we briefly discuss the background to GP and details about the kernels we use later on. We also discuss our implementation of GP using cosmological data sets at background level. Sec.~\ref{sec:gp_abc_smc} then delves into the main analysis of the work where we first explore the use of ABC-SMC for both kinds of data sets being considered here. We then apply GAs in order to determine the best acting GP kernel in a mutating population of kernels. Finally, we summarize our core results in Sec.~\ref{sec:conclusions} and discuss possible future work on the topic. 

We implement our calculations in \textit{python} through \textit{jupyter notebooks} \cite{jupyter} and acknowledge the use of the packages \textit{pyabc} \cite{10.1093/bioinformatics/bty361}, \textit{pygad} \cite{pygad}, and \textit{geneal} \cite{geneal}. The reader interested in recreating our output and exploring further is highly encouraged to use our  jupyter notebooks \cite{reggie_bernardo_4810864}.

\section{Gaussian Processes in Late-Time Cosmology}
\label{sec:gaussian_processes}
In this section, we provide a brief introduction to Gaussian processes (Sec.~\ref{subsec:gp_recap}) and its application to late-time cosmological data sets (Sec.~\ref{subsec:cosmology_gp}).

\subsection{Gaussian processes}
\label{subsec:gp_recap}

The GP regression is an emerging go-to cosmology-independent tool that exploits the use of a kernel to make predictions on observational parameters \cite{10.5555/971143, 10.5555/1162254}. In light of the existing tensions between early (i.e. during last scattering), and local cosmological observations, GP has naturally become popular as a refreshing change of view in making cosmological predictions that have often been based on arbitrary parametrizations of the underlying theory \cite{Seikel2012, Seikel:2013fda, Shafieloo:2012ht, Colgain:2021ngq,Yennapureddy:2017vvb, Benisty:2020kdt, Belgacem:2019zzu,Moore:2015sza,Canas-Herrera:2021qxs, Briffa:2020qli,Cai:2019bdh, LeviSaid:2021yat, Cai:2015zoa, Reyes:2021owe, Wang:2017jdm, Gomez-Valent:2018hwc, Zhang:2018gjb, Mukherjee:2020vkx, Aljaf:2020eqh, Li:2019nux, Liao:2019qoc, Busti:2014aoa, Cai:2015pia}.

Consider an observation of $N$ data points $\left( z, H(z) \right)$ with uncertainties contained in a covariance matrix $C$. To reconstruct the function $H(z^*)$ at the coordinates $z^*$, GP relies on a kernel $K\left( z^* , \tilde{z}^* \right)$, or a covariance function, to connect the function values at coordinates $z^*$ and $\tilde{z}^* \neq z^*$. In terms of the kernel, the mean and the covariance of the GP reconstruction of the $n$th derivative of $H(z)$ at $z^*$ are given by
\begin{equation}
\label{eq:gp_ave}
    \langle H^{* (n)} \rangle = K^{(n, 0)} \left( z^*, Z \right) \left[ K\left( Z, Z \right) + C \right]^{-1} H \left( Z \right)\,,
\end{equation}
and
\begin{equation}
\label{eq:gp_cov}
    \text{cov} \left( H^{* (n)} \right) = K^{(n, n)} \left( z^*, z^* \right) - K^{(n, 0)} \left( z^*, Z \right) \left[ K\left(Z, Z\right) + C \right]^{-1} K^{(0, n)} \left(Z, z^*\right)\,,
\end{equation}
respectively, where $Z$ stands for the union of the redshifts of the measurements and $y^{(n, m)}$ refers to the $n$th derivative of a function $y$ with respect to its first argument and the $m$th derivative with respect to the second argument. Now, it should be emphasized that the kernel depends on a set of hyperparameters $\theta$ that will be trained to describe the characteristics of the particular data sets under consideration. To be more precise, the hyperparameters $\theta$ are determined by marginalizing over the marginal likelihood $\mathcal{L} = p \left( H | Z, \theta \right)$ where
\begin{equation}
\label{eq:logmlike}
\ln \mathcal{L} = -\dfrac{1}{2} H\left(Z\right)^{T} \left[ K\left(Z, Z\right) + C \right]^{-1} H\left(Z\right) - \dfrac{1}{2} \ln | K\left(Z, Z\right) + C | - \dfrac{N}{2} \ln \left(2\pi\right)\,.
\end{equation}
Eqs.~(\ref{eq:gp_ave}--\ref{eq:logmlike}) flesh out the GP methodology which can readily be implemented. The simplicity of this formula makes GP regression, as a modelling tool, very direct to utilize in conjunction with other strategies.

Despite these advantageous properties, GP is anchored by the choice of the kernel $K \left( z^*, \tilde{z}^* \right)$. However, in applications, the choice of kernel can often be determined in seemingly arbitrary ways, without prior knowledge of machine learning or the underlying theory to reconstruct. When confronted with such uneasy decision, then the understandable way to go is to obtain results coming from \textit{all} imaginable kernels instead of making a single choice. Indeed, this pragmatic approach has been adopted in the cosmology community and with it, outstanding results have been obtained, supporting its practicality in these applications. However, the question of which kernel is the best applicable one remains an open question. This will be the explored in this work using the vehicle of expansion data as a reference point.

As a base for the kernel selection problem, we consider three of the most widely used kernels for GP in the literature, namely, the radial basis function, rational quadratic, and the Mat\'{e}rn kernels. Their functional forms are listed below:
\begin{itemize}
\item Radial basis function (RBF)
\begin{equation}
\label{eq:rbf}
K(r) = A^2 \exp \left( - \dfrac{ r^2 }{2l^2} \right)\,,
\end{equation}
also often referred to as the \textit{squared exponential kernel}. This kernel is infinitely differentiable and so can be used in a GP to reconstruct a function and any of its derivatives. 
\item Rational quadratic (RQ)
\begin{equation}
\label{eq:rq}
K(r) = A^2 \left( 1 + \dfrac{ r^2 }{2 \alpha l^2} \right)^{-\alpha}\,.
\end{equation}
A special case of RQ is the Cauchy kernel (CHY) for $\alpha = 1$. Like the RBF kernel, the RQ kernel is infinitely differentiable and can be used to reconstruct a function and any of its derivatives.
\item Mat\'{e}rn with index $\nu = x/2$ (M$x$2)
\begin{equation}
\label{eq:matern}
K(r) = A^2 \dfrac{2^{1 - \nu}}{\Gamma \left( \nu \right)} \left( \dfrac{\sqrt{2\nu r^2}}{l} \right)^\nu K_\nu \left( \dfrac{\sqrt{2\nu r^2}}{l} \right)\,,
\end{equation}
where $K_\nu \left(x \right)$ is the modified Bessel function and $\nu$ is a positive constant. The Mat\'{e}rn kernel can be differentiated $n$ times with $n < \nu$. Appealing choices in physics applications are therefore the at least-twice-differentiable Mat\'{e}rn kernels $\nu = 5/2$ (M52) and $\nu = 7/2$ (M72).
\end{itemize}
We shall use the kernels above both independently and in a hybrid  fashion in order to reconstruct cosmological data sets.

\subsection{Cosmology through Gaussian processes}
\label{subsec:cosmology_gp}

We use two types of expansion data, cosmic chronometers (CC) and the supernova type 1a Pantheon data set (SNe). The CC data set comprises of points mainly within the $z \lesssim 2$ range and produces $H(z)$ data without relying on a cosmological model \cite{Moresco:2016mzx, Moresco:2015cya, 2014RAA....14.1221Z, 2010JCAP...02..008S, 2012JCAP...08..006M,Ratsimbazafy:2017vga}. This depends on a differential aging technique between galaxies. For the SNe data, we utilize the full Pantheon data set \cite{Scolnic:2017caz} which describes expansion through the distance modulus $m(z)$ and consists of 1024 points. Here, Cepheids are used to calibrate distance measurements for SNe events.

\begin{figure}[ht]
\center
	\subfigure[ ]{
		\includegraphics[width = 0.47 \textwidth]{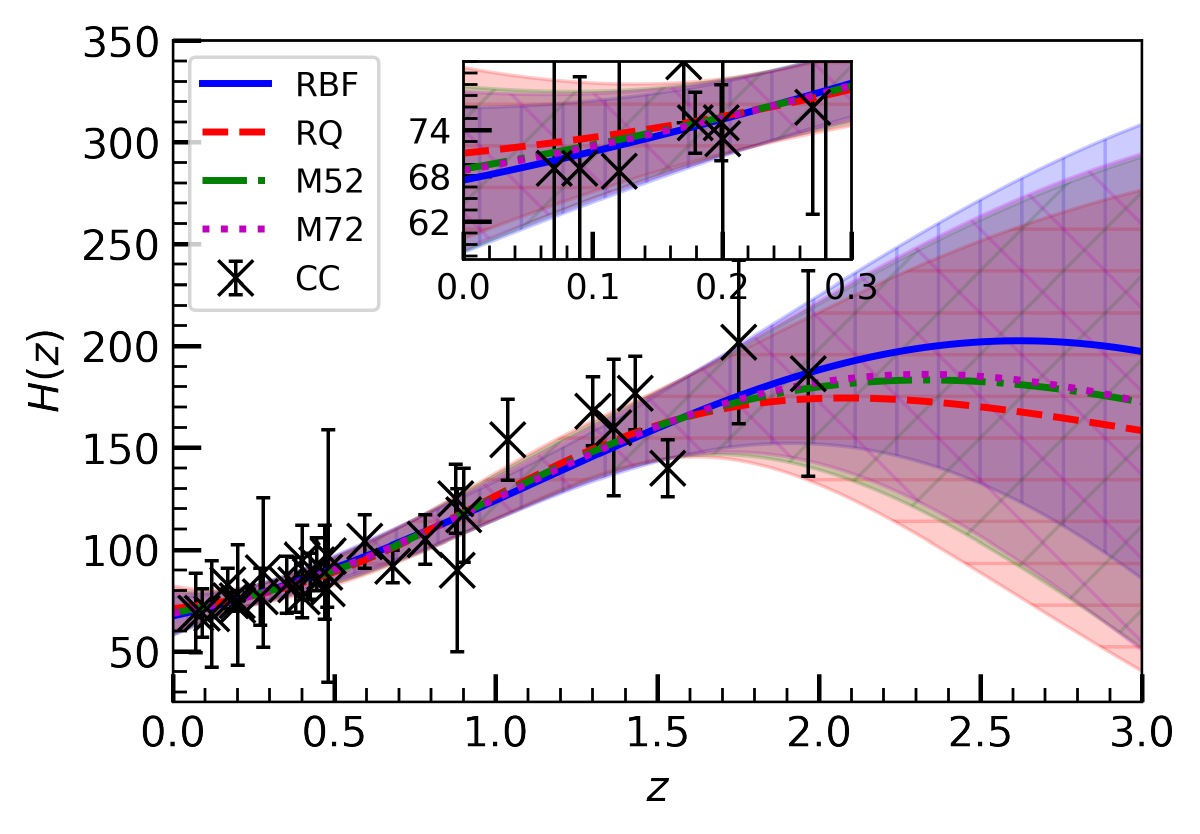}
		}
	\subfigure[ ]{
		\includegraphics[width = 0.47 \textwidth]{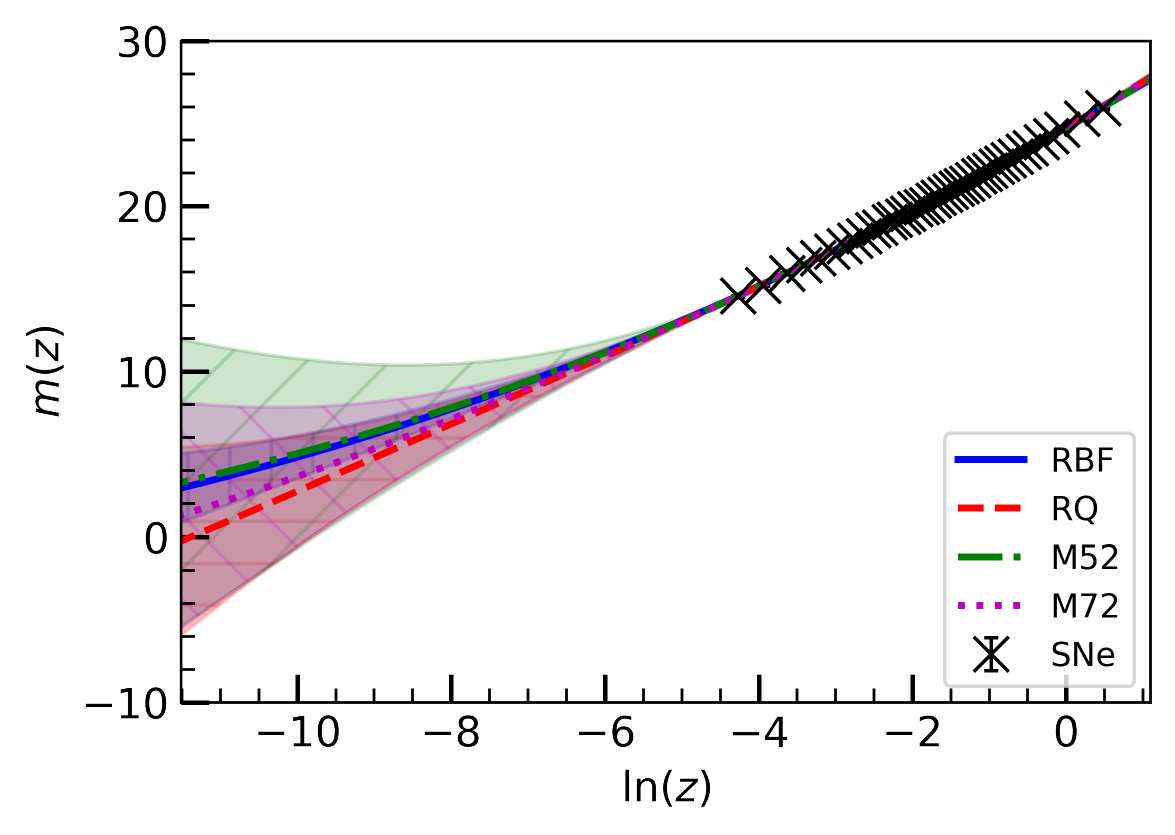}
		}
\caption{GP reconstructions of the (a) CC and (b) SNe data sets obtained by optimizing over the log-marginal likelihood (Eq. (\ref{eq:logmlike})). The colored-hatched regions show the $2\sigma$ region around the mean of the GP. Hatches used: (RBF, $'|'$), (RQ, $'-'$), (M52, $'/'$), (M72, $' \text{\textbackslash} '$).}
\label{fig:H0_kernel_tension}
\end{figure}

In Fig.~\ref{fig:H0_kernel_tension} we reconstruct both Hubble function using CC data and the distance modulus using the Pantheon data set using vanilla GP regression. As expected the CC reconstruction is strong for low redshifts and starts to increase in uncertainties for higher redshifts, while the SNe GP reconstruction has very low uncertainties throughout the reconstructed region. In fact, we show the Pantheon data set against $\log z$ to highlight the slight differences near the origin. It is also important to point out that, by and large, the kernels agree to within 1$\sigma$ uncertainties on their predicted mean values for the respective reconstructed functions. However, the slight differences due emerge from the intrinsic differences between the kernel functions themselves and so we need a strategy to understand better which of the kernels perform better for different data sets.

In the same vein, the choice of the kernel becomes important outside of the boundaries of the observation. In Fig.~\ref{fig:H0_kernel_tension} (a), this appears at $z = 0$ and $z > 2$. At $z = 0$, the corresponding predictions of the Hubble parameter today are presented in Table \ref{tab:H0_kernel_tension}. This shows that the prediction based on RQ, the most general of the kernels considerd here, is in 0.8$\sigma$ tension with the RBF prediction. Clearly, with $H_0$ being arguably the most important quantity in the late-time Universe, the choice of a kernel should be taken more seriously. For $z > 2$, similar sentiments can be drawn. At $z \sim 3$, the RQ mean can already be seen to be near the $1\sigma$ edges of the RBF prediction. In Fig.~\ref{fig:H0_kernel_tension} (b), for the reconstructed SNe distance modulus, it can be seen that tensions between the different kernel-based predictions at $\ln(z) < -5$ also appears and arguably even more severe. In this case, the M72 prediction is already hovering near the $2\sigma$ boundary of the RBF prediction while the RQ prediction is completely outside it. There can of course also be agreement between the predictions of two different kernels. Such a case is shown in Fig.~\ref{fig:H0_kernel_tension} (a) for the two Mat\'{e}rn predictions, M52 and M72, and in Fig.~\ref{fig:H0_kernel_tension} (b) for the RBF and M52 predictions.

To set a baseline with respect to parametric reconstructions, it is also useful to refer to the $\chi^2$ value given by
\begin{equation}
\label{eq:chi2}
\chi^2 = \left( E \left[ H(Z) \right] - H (Z) \right)^T C^{-1} \left( E \left[ H(Z) \right] - H (Z) \right)\,,
\end{equation}
where $E \left[ H(Z) \right]$ is the mean of the GP reconstructed function $H(z^*)$ at the data positions $z^* = Z$ and $C^{-1}$ is the matrix inverse of the data covariance matrix. This quantifies the performance of the reconstruction against the original observational data. For the optimization-based GP reconstruction of $H(z)$ (Fig.~\ref{fig:H0_kernel_tension} (a)), the log-marginal likelihood and $\chi^2$ are shown in Table \ref{tab:Hz_rec_gp_opt}.

\begin{table}[h!]
\center
\caption{Log-marginal likelihood and $\chi^2$ for the reconstructed $H(z)$ function in Fig.~\ref{fig:H0_kernel_tension} (a).}
\begin{tabular}{| c | c | c |}
\hline \hline
$\phantom{ \dfrac{1}{1} }$ kernel $\phantom{ \dfrac{1}{1} }$ & $\ln \mathcal{L}$ & $\chi^2$ \\ \hline
RBF & $-131.85$ & 14.3 \\ \hline
RQ  & $-131.47$ & 12.9 \\ \hline
M52  & $-131.56$ & 13.3 \\ \hline
M72 & $-131.65$ & 13.6 \\ \hline
\end{tabular}
\label{tab:Hz_rec_gp_opt}
\end{table}

The corresponding statistics for the reconstructed $m(z)$ function (Fig.~\ref{fig:H0_kernel_tension} (b)) are presented in Table \ref{tab:mz_rec_gp_opt}. One source of the drastic difference in values is that the Pantheon data set contains 1024 points.

\begin{table}[h!]
\center
\caption{Log-marginal likelihood and $\chi^2$ for the reconstructed $m(z)$ function in Fig.~\ref{fig:H0_kernel_tension} (b).}
\begin{tabular}{| c | c | c |}
\hline \hline
$\phantom{ \dfrac{1}{1} }$ kernel $\phantom{ \dfrac{1}{1} }$ & $\ln \mathcal{L}$ & $\chi^2$ \\ \hline
RBF & $60.2$ & $45.1$ \\ \hline
RQ  & $60.7$ & $41.1$ \\ \hline
M52  & $62.3$ & $34.4$ \\ \hline
M72 & $61.2$ & $40.1$ \\ \hline
\end{tabular}
\label{tab:mz_rec_gp_opt}
\end{table}

For the CC data set, $\chi^2 < N = 31$ for all of the reconstructions, reflective of overfitting, which is a common pathology shared by non-parametric reconstruction methods. Nonetheless, for the binned Pantheon data set, the overfitting did not seem to manifest as strongly, except for M52 where $\chi^2 < N = 40$. It should also be pointed out that Tables \ref{tab:Hz_rec_gp_opt} and \ref{tab:mz_rec_gp_opt} seem to suggest a relation between the log-marginal likelihood and the $\chi^2$ value. Our results indeed support the assertion that maximizing the log-marginal likelihood almost always leads to smaller $\chi^2$ values. However, we shall caution that the inverse does not follow, i.e., minimizing the $\chi^2$ value do not correspond to maximizing the log-marginal likelihood.

In the sections that follow, we will confront the question of whether any one of these kernels is preferred over the others using evolutionary inspire algorithms.

\section{Gaussian Processes in an Approximate Bayesian Computation--Sequential Monte Carlo Framework}
\label{sec:gp_abc_smc}

Using the ABC-SMC method, we here explore how this can be employed to resolve the kernel tension problem for both the CC and SNe data sets.

\subsection{Approximate Bayesian computation with sequential Monte Carlo sampling}
\label{sec:abc_smc}

ABC is an inference algorithm that can be used to estimate the parameters of a model even when the likelihoods are unknown. It is based on the concept that the likelihood can be sufficiently approximated when the distance between a sampled set of predictions and the observations is below some tolerance value. A more sophisticated version of this algorithm, adapting a SMC sampling approach, leads to a more powerful tool which can make inevitably improving predictions per sampling generation. The resulting ABC-SMC method \cite{2009arXiv0911.1705T} has been successfully applied in a wide variety of problems including astrophysics and cosmology \cite{Akeret:2015uha, 2013ApJ...764..116W, 2017A&C....19...16J, Ishida:2015wla} and, in line with the topic remit of this work, namely for the kernel selection problem \cite{10.3389/fbuil.2017.00052}. We summarize the main steps of this algorithm in what follows.

The goal of ABC is to be able to approximate the posterior distribution 
\begin{equation}
    P \left( \theta | D , \mathcal{M} \right) \propto \mathcal{L} \left( D | \theta, \mathcal{M} \right)\,,
\end{equation}
where $\theta$ is a set of parameters describing the model $\mathcal{M}$, $D$ is the data, and $\mathcal{L} \left( D | \theta, \mathcal{M} \right)$ is the likelihood of the data to be represented by the parameters $\theta$ and model $\mathcal{M}$. To be able to do this without a likelihood, and in a computationally-efficient manner, the ABC-SMC relies instead on a distance function $\Delta(R)$ to measure how far away a certain prediction is from the data. Typical choices in ABC applications are the absolute distance and the mean-squared error. On the other hand, for the GP application, it is the the log-marginal likelihood (Eq. (\ref{eq:logmlike})) that is the natural choice for the distance function. The tolerance $\varepsilon$ guaranteeing the acceptance of a randomly drawn \textit{particle} $\left( \mathcal{M}, \theta \right)$ if and only if $\Delta(R) \leq \varepsilon$ is then sequentially-tightened with each \textit{population}. The crucial ingredient which makes the ABC-SMC computationally-efficient is that the samples of the newer populations are always drawn from the previous one. In this case, the newer populations can only continue to improve on its predecessor and always come with a better, more reliable, approximation of both the model and the parameter posterior distributions. We refer the reader to Ref. \cite{2009arXiv0911.1705T} for an excellent introduction to ABC-SMC and also to the python package \textit{pyabc}  \cite{10.1093/bioinformatics/bty361}.

In the following, we will apply the ABC-SMC method to the kernel selection problem in GP. A similar study was performed in Ref.~\cite{10.3389/fbuil.2017.00052} but the particulars of real data was not implemented in the previous study.

\subsection{GP-ABC-SMC Implemented on CC and SNe Late-Time Data}
\label{subsec:gp_abc_smc_cosmology}

We consider for this section the two-hyperparameter family of kernels, i.e., those described only by a pair of length scale $l$ and amplitude height $A$ hyperparameters. These are the RBF, CHY, and Mat\'{e}rn (M52 \& M72) kernels. The posterior distribution on the joint space of these kernels will then be obtained through the ABC-SMC beginning (at generation $t = -1$) with a uniform prior distribution on both the kernel and hyperparameter spaces. In our implementation, we use an adaptive strategy that automatically determines the tolerances $\varepsilon$ per population and the size of the population for both CC and SNe applications. Most importantly, we consider the GP's log-marginal likelihood (Eq.~(\ref{eq:logmlike})) as the distance function for the ABC pieces of the overall algorithm.

\begin{figure}[ht]
\center
	\subfigure[ ]{
		\includegraphics[width = 0.47 \textwidth]{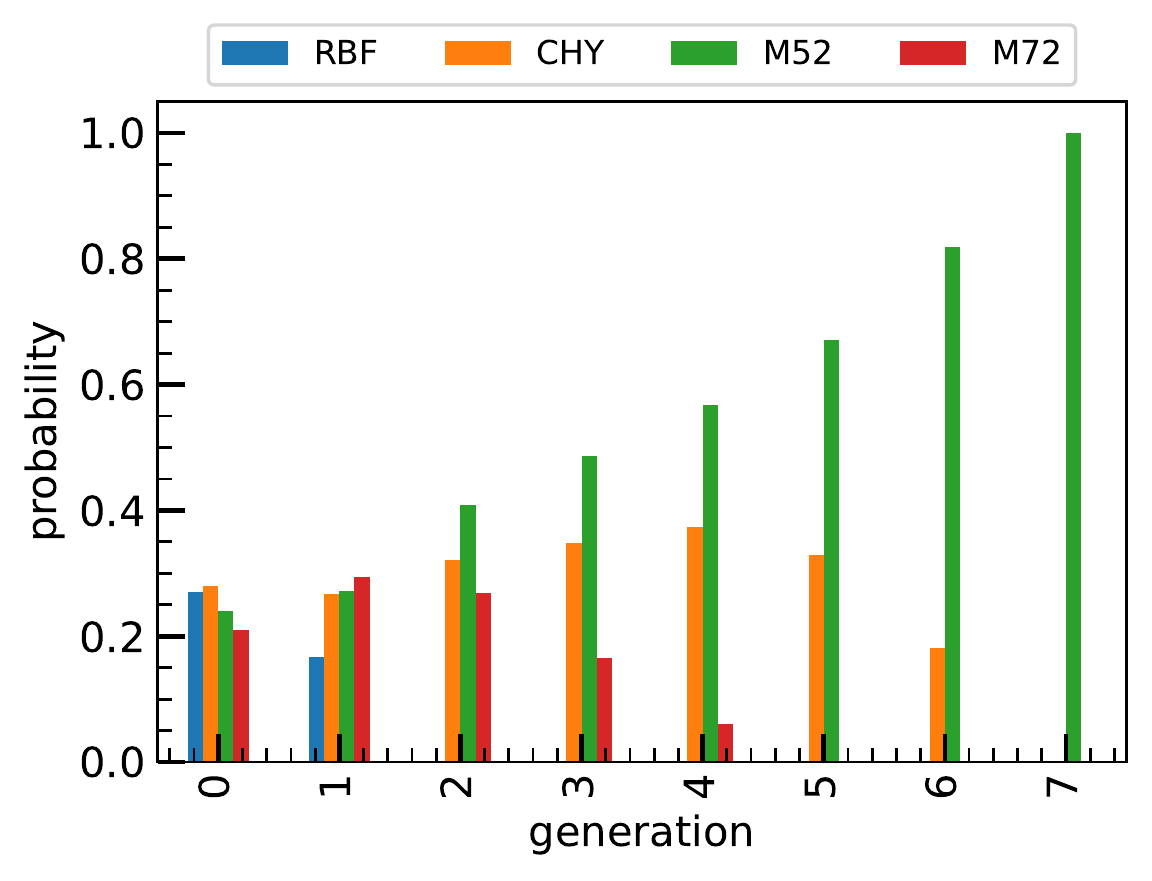}
		}
	\subfigure[ ]{
		\includegraphics[width = 0.47 \textwidth]{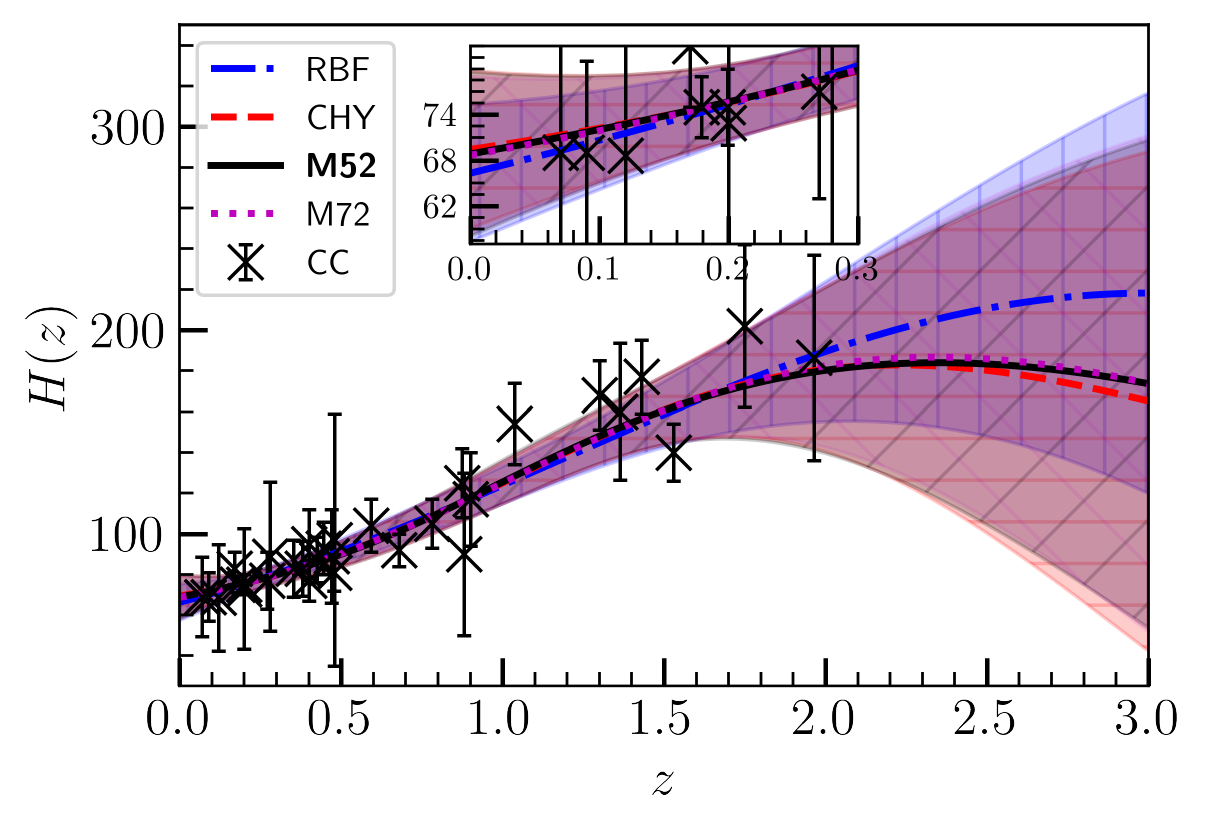}
		}
	\subfigure[ ]{
		\includegraphics[width = 0.47 \textwidth]{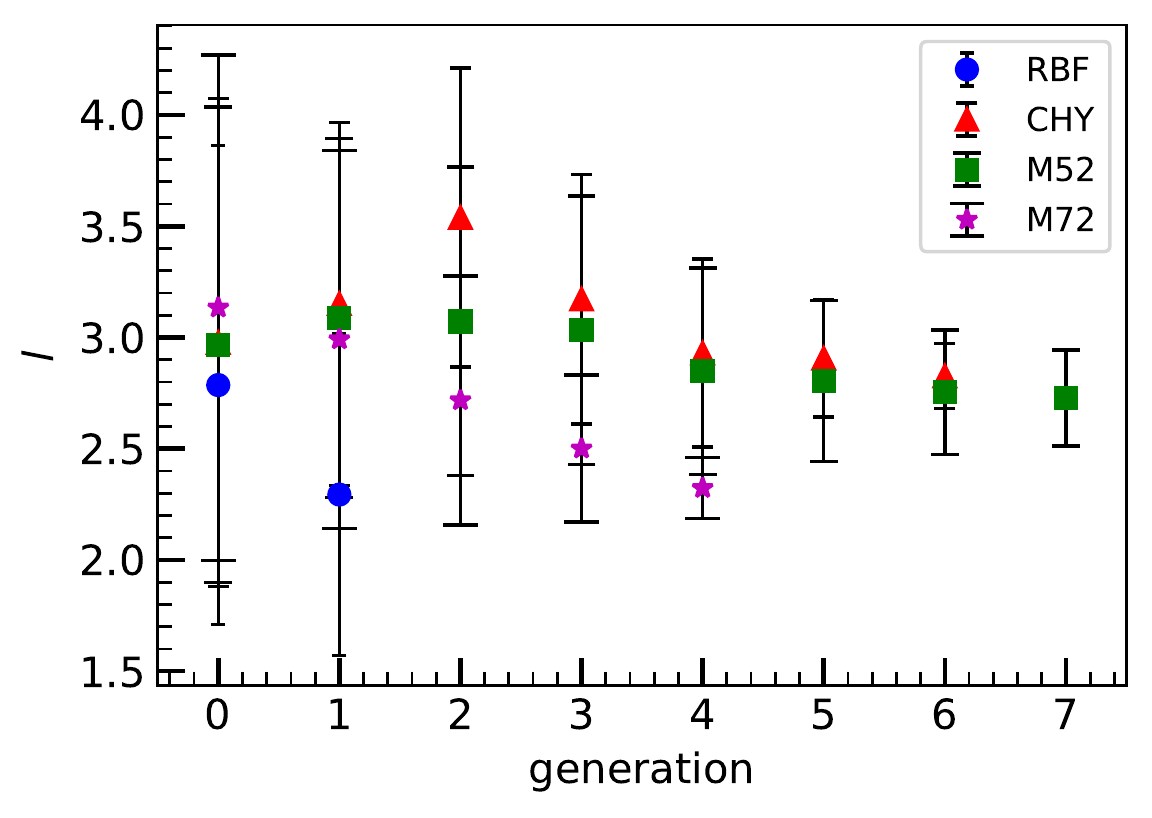}
		}
	\subfigure[ ]{
		\includegraphics[width = 0.47 \textwidth]{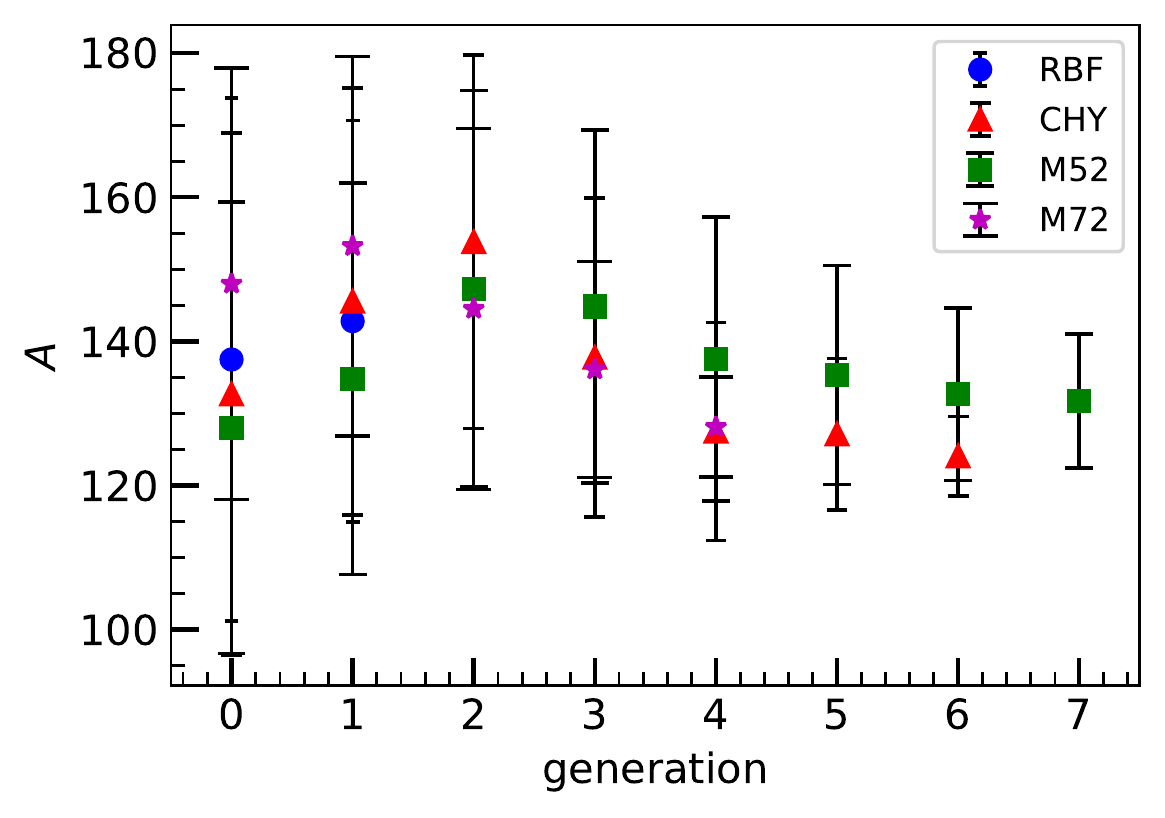}
		}
\caption{The output of a GP-ABC-SMC run for the $H(z)$ reconstruction of the CC data set: (a) Joint kernel posterior (b) prediction of the last surviving population per kernel (c) posterior estimate of $l$ per generation (d) posterior estimate of $A$ per generation. The colored-hatched regions in (b) show the $2\sigma$ region around the mean. Hatches used: (RBF, $'|'$), (RQ, $'-'$), (M52, $'/'$), (M72, $' \text{\textbackslash} '$). The error bars in (c) and (d) show the $1\sigma$ region of the hyperparameter posterior.}
\label{fig:CC_abcsmc}
\end{figure}

The result of the combined GP and ABC-SMC run for the CC data set is shown in Fig.~\ref{fig:CC_abcsmc}. It can be observed that the M52 kernel is singled out to be the preferable GP kernel after just eight generations. We emphasize that this conclusion has been observed in all the GP-ABC-SMC runs performed with varying population sizes and strategies. Fig.~\ref{fig:CC_abcsmc} (a) represents a sample of this run that lead to the same inevitable result. Fig.~\ref{fig:CC_abcsmc} (b) shows the prediction of the last surviving population per kernel. Interestingly, the prediction based on the RBF kernel, which only survived for two generations, already appear to be a visually-acceptable GP where the data points are clustered. Also, the mean and the uncertainty of the GP based on CHY, M52, and M72 kernels remained close with each other despite the populations of M72 and CHY dying at different stages throughout generations. 

This demonstrates, if nothing else, the power of the GP-ABC-SMC in singling out a kernel in reconstructing an observational data set. Moreover, in the spirit of Bayesian analysis, Figs.~\ref{fig:CC_abcsmc} (c) and \ref{fig:CC_abcsmc} (d) show the evolution of the hyperparameter posteriors throughout each generation. Indeed, with each passing generation, it can be seen that the hyperparameter posterior for any one of the kernels becomes a narrower, better, approximation of the true posterior. The GP-ABC-SMC method implemented here therefore clearly not just selects the preferential kernel for a given data set but also already obtains the hyperparameters of the surviving kernel in the process.

\begin{figure}[ht]
\center
	\subfigure[ ]{
		\includegraphics[width = 0.47 \textwidth]{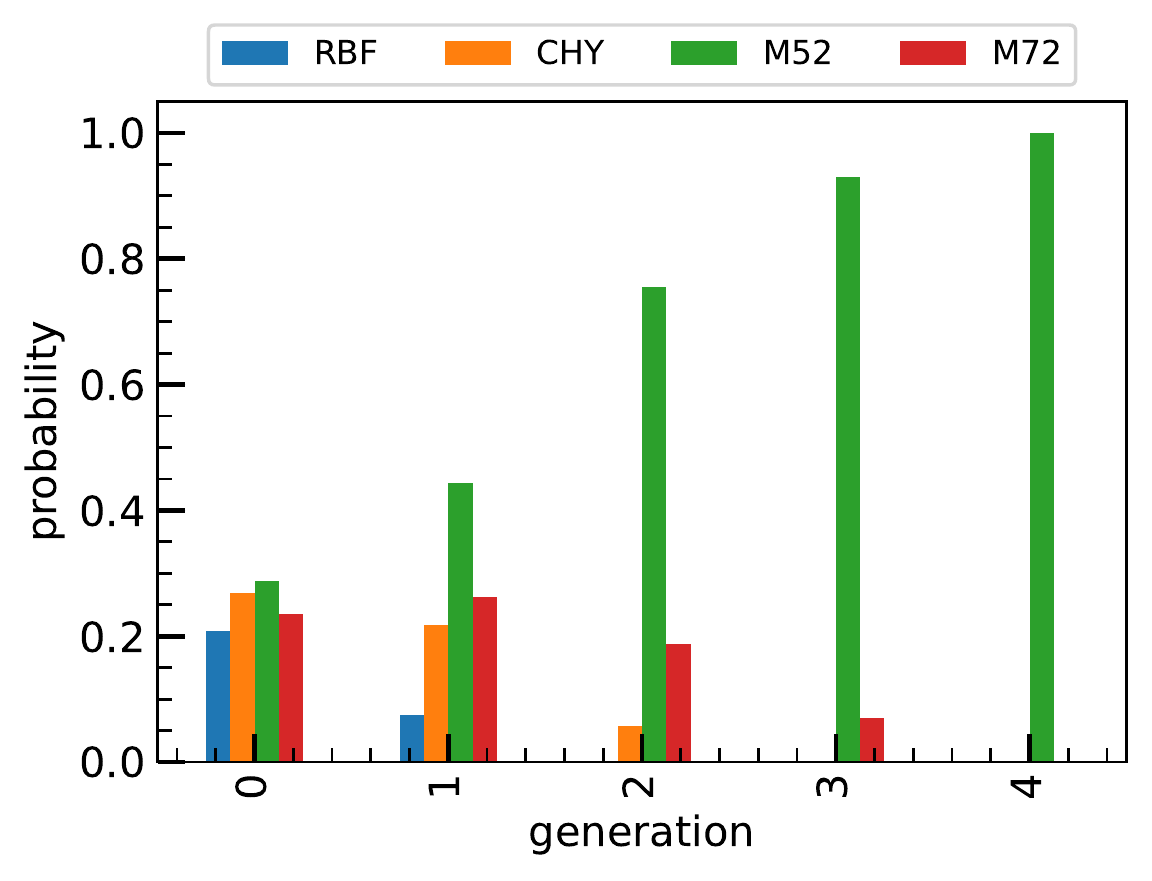}
		}
	\subfigure[ ]{
		\includegraphics[width = 0.47 \textwidth]{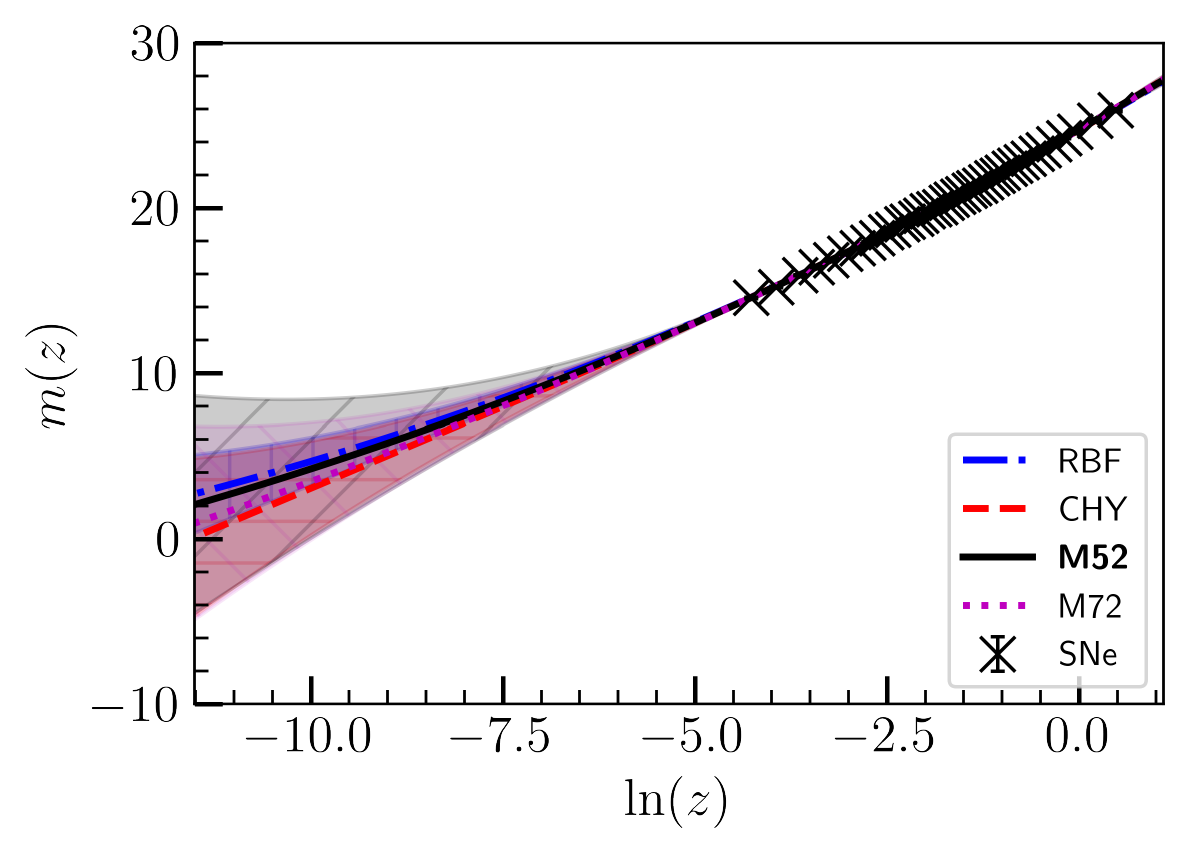}
		}
	\subfigure[ ]{
		\includegraphics[width = 0.47 \textwidth]{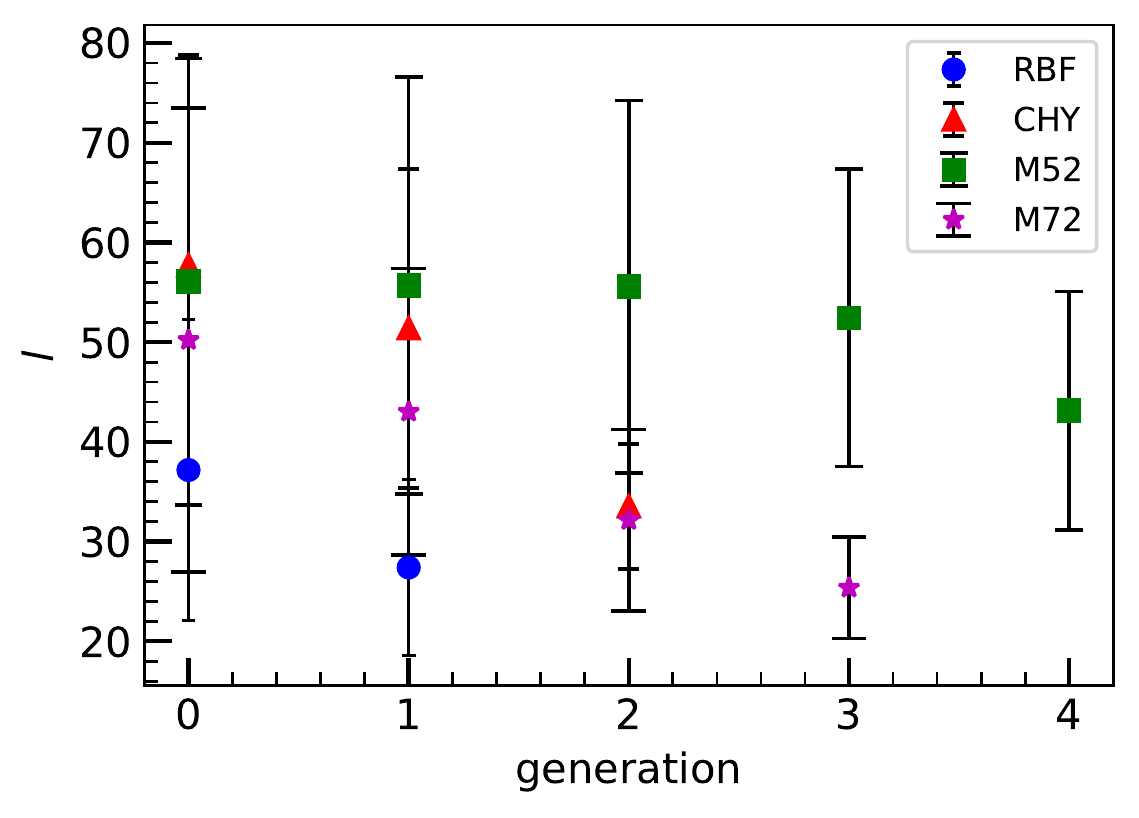}
		}
	\subfigure[ ]{
		\includegraphics[width = 0.47 \textwidth]{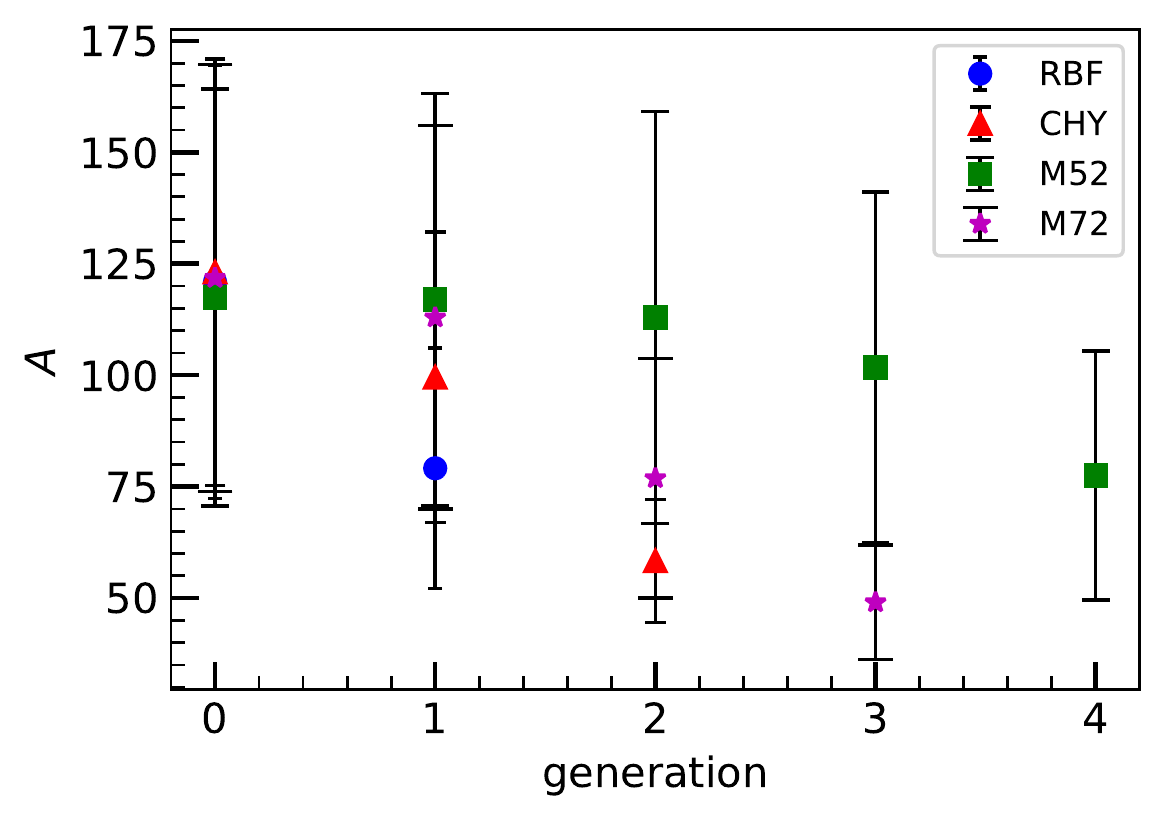}
		}
\caption{The output of a GP-ABC-SMC run for the $m(z)$ reconstruction of the SNe data set: (a) Joint kernel posterior (b) prediction of the last surviving population per kernel (c) posterior estimate of $l$ per generation (d) posterior estimate of $A$ per generation. The colored-hatched regions in (b) show the $2\sigma$ region around the mean. Hatches used: (RBF, $'|'$), (RQ, $'-'$), (M52, $'/'$), (M72, $' \text{\textbackslash} '$). The error bars in (c) and (d) show the $1\sigma$ region of the hyperparameter posterior.}
\label{fig:SN_abcsmc}
\end{figure}

The result of the GP-ABC-SMC analysis on the Pantheon SNe data set is shown in Fig.~\ref{fig:SN_abcsmc}. It is most interesting that the M52 kernel was singled out again as the more preferable among the two-hyperparameter family of kernels (Fig.~\ref{fig:SN_abcsmc} (a)). This conclusion should be given due attention since the SNe data is a lot more stringent, very small uncertainties, and notably comes with a full covariance matrix, i.e., with nonzero off-diagonal elements, that was used in the ABC-SMC runs.

The prediction of the last surviving population for each kernel is shown in Fig.~\ref{fig:SN_abcsmc} (b). Once again, we find that the predictions of any one kernel, even those coming from dying populations, can be visually-accepted as a reconstruction of the data set in places where the observations are clustered, in this case, for $\ln (z) \gtrsim - 5$. At points far away from the light of observations, $\ln (z) < - 5$, we find that the predictions more or less resonates also with their optimization-based results (Fig.~\ref{fig:H0_kernel_tension}), i.e., the RBF turns out to have the smallest uncertainty and the CHY and M72 means are in a nearly-$2\sigma$ tension with the RBF prediction. It is worth pointing out that the uncertainties emerging from the GP-ABC-SMC run turn out to be smaller than their optimization-counterparts, which happens due to the optimization that is taking place in the ABC-SMC part of the algorithm (see Sec.~\ref{sec:abc_smc}). We clarify that the GP reconstructions take only the mean hyperparameters of the last surviving population per kernel. The evolution of the hyperparameter posterior for this SNe application is shown in Figs.~\ref{fig:SN_abcsmc} (c) and \ref{fig:SN_abcsmc} (d). The hyperparameter posteriors for each kernel always become inevitably narrower with each passing generation and eventually converges to the true posterior.

\section{Genetic Algorithm Kernel Selection: Implementation and Results}
\label{sec:gp_ga}

We introduce the genetic algorithm (Sec.~\ref{subsec:ga}) and use it to determine a most naturally-fit kernel from a competitive pool of a ten hyperparameter class of kernels (Sec.~\ref{subsec:gp_ga_cosmology}).

\subsection{Genetic algorithm}
\label{subsec:ga}

GA is a collection of optimization tools that inherits properties from the process of natural selection in evolution theory. Like the ABC-SMC method, GA samples from a pool of individuals, making up a population which are then ranked according to their fitness. The fittest individuals are then given better chances to proceed to the next generation. Also, they carry on to also make offsprings to replace the ones left out by the natural selection. These surviving individuals are given a chance to mutate. The crossover and mutation mechanisms are carried out at the level of the genes making up the chromosomes uniquely characterizing an individual. In this way, fitter kernels will be allowed to survive and mutate to further span the space of kernels that show promise in approximating the data sets.

GA depends on the following key ingredients:
\begin{itemize}
\item \textit{Fitness function}: Describes the order by which the individuals of an existing population will be ranked. In optimization, this can be taken to be the function to be maximized. In this light, the fitness function scores the solutions in a particular generation using pre-determined rubric;
\item \textit{Selection}: Denotes the fraction of the population that will be chosen to proceed to the next generation. Typically, a ``roulette wheel'' system is implemented whereas the fittest individuals are given better chances of survival;
\item \textit{Crossover}: Describes the mating of parents, mixing their genes, to produce offsprings that would takeover the place of the individuals left out in the selection process. Crossover is then the process by which pairs of surviving elements of a population are combined to produce new elements that may be fitter in terms of the fitness function. In the regime of kernels, this would mean that kernel functions would combine to produce more complex kernels, which may perform better in terms of the fitness function but which will at some point become disfavoured due to their increase in complexity;
\item \textit{Mutation}: Describes the particular way in which an individual mutates, or rather, is genetically-altered. This important step in the GA is key to producing a stronger, fitter, individuals. Thus, these are small changes in the elements of a population which may produce elements in the next generation that are fitter. In the context of the kernel selection problem this may, for instance, add or multiply a small portion of kernels in the next generation by a kernel from a base alphabet of kernels.
\end{itemize}

We refer the reader to the python package \textit{pygad} \cite{pygad} for more details and an awesome introduction to GA with many illustrative examples. One of the main advantages of the GA in optimization is that it does not scale badly with the number of parameters. This makes it the ideal tool when we apply it to a kernel with \textit{ten} hyperparameters. On the other hand, certain Bayesian approaches such as MCMC suffer greatly in these scenarios.

In the results, we use GA to single out a particular set of hyperparameters for the GP reconstruction with the kernel
\begin{equation}
\label{eq:individual}
K(r| \theta) = C_{\text{RBF}}^2 K_\text{RBF}( r | l_{\text{RBF}})^{n_{\text{RBF}}} + C_{\text{RQ}}^2 K_\text{RQ}( r | l_{\text{RQ}}, \alpha_{\text{RQ}})^{n_{\text{RQ}}} + C_{\text{M52}}^2 K_\text{M52}( r | l_{\text{M52}})^{n_{\text{M52}}} \,,
\end{equation}
where
\begin{equation}
\label{eq:k_rbf}
K_{\text{RBF}}( r | l) = \exp \left( - \dfrac{ r^2 }{2l^2} \right)\,,
\end{equation}
\begin{equation}
\label{eq:k_rq}
K_\text{RQ}( r | l, \alpha) = \left( 1 + \dfrac{ r^2 }{2 \alpha l^2} \right)^{-\alpha}\,,
\end{equation}
and
\begin{equation}
\label{eq:k_m52}
K_{\text{M52}}( r | l) = \exp \left( - \dfrac{\sqrt{5} r}{l} \right) \left( 1 + \dfrac{\sqrt{5} r}{l} + \dfrac{5 r^2}{3 l^2} \right) \,.
\end{equation}
Obviously, Eq.~(\ref{eq:individual}) is a mixture of the RBF, RQ, and M52 \textit{subkernels} and is determined by a total of ten hyperparameters. The goal of this section is to use the GA to find the optimal set of hyperparameters for the GP reconstruction of our Hubble expansion data.

In the language of GA, the kernels themselves (\ref{eq:individual}) are the individuals of a population and the chromosomes are the set of hyperparameters $\theta$ describing each kernel. Thus, the chromosome will be described by
\begin{equation}
\label{eq:chromosome}
\theta = \left( C_{\text{RBF}}, l_{\text{RBF}} , n_{\text{RBF}}, C_{\text{RQ}}, l_{\text{RQ}}, \alpha_{\text{RQ}}, n_{\text{RQ}}, C_{\text{M52}}, l_{\text{M52}} , n_{\text{M52}} \right)\,,
\end{equation}
where the hyperparameters are the genes characterizing a chromosome.

Now, to measure the fitness of a kernel, we consider a Bayesian information criterion (BIC)-inspired quantity
\begin{equation}
\label{eq:fitness_function}
    \mathcal{F} = \ln \mathcal{L} - \dfrac{k_\text{eff} \ln N}{2}\,,
\end{equation}
where $\ln \mathcal{L}$ is the log-marginal likelihood Eq.~(\ref{eq:logmlike}) and $k_\text{eff}$ is the effective number of hyperparameters describing the kernel. The term $k_\text{eff} \ln N /2$ acts as a penalty for more complex kernels and is determined on the following grounds: if the product $l_i C_i > \eta$ for some constant $\eta$ where $i = $ RBF, RQ, and/or M52, then the number of hyperparameters of the subkernel $i$ is added to $k_\text{eff}$. For example, if $l_\text{RBF} C_\text{RBF} > \eta$ and $l_\text{RQ} C_\text{RQ} > \eta$, then $k_\text{eff} = 3 + 4 = 7$. Therefore, generally, the majority of the individuals in a diversified population will have $k_\text{eff} = 10$ and so must pay a larger penalty to justify their complexity. We set $\eta$ to $\eta = 10^{-3}$ in the trials to be discussed. This means that a GA will quickly tend to a fittest kernel rather produce extremely large hyperparameter kernels that take many generations to settle to a final evolutionary state.

\subsection{GP-GA Implemented on CC and SNe Late-Time Data}
\label{subsec:gp_ga_cosmology}
 
The results have shown that the mutation plays a major role in the optimization process. We first discuss the results on the CC data set reconstruction and then finish with the SNe data set. 

Table \ref{tab:GA_runs} shows the parameters and the fitness of the best individual for several trials with the GA on the CC data set.
\begin{table}[h!]
\center
\caption{Parameters and best fitness for four trials with the GA with $n_\text{gen}$ generations for the CC data set reconstruction.}
\begin{tabular}{| c | c | c | c | c | c |}
\hline
trial & population size & selection rate & mutation rate & $n_\text{gen}$ & best fitness \\ \hline
$\phantom{\dfrac{1}{1}}$ 1 $\phantom{\dfrac{1}{1}}$ & $10^4$ & 0.5 & 0.15 & $10^1$ & $-143.5$  \\ \hline
$\phantom{\dfrac{1}{1}}$ 2 $\phantom{\dfrac{1}{1}}$ & $10^4$ & 0.3 & 0.30 & $10^1$ & $-148.5$ \\ \hline
$\phantom{\dfrac{1}{1}}$ 3 $\phantom{\dfrac{1}{1}}$ & $10^3$ & 0.1 & 0.10 & $10^2$ & $-143.4$ \\ \hline
$\phantom{\dfrac{1}{1}}$ 4 $\phantom{\dfrac{1}{1}}$ & $10^3$ & 0.3 & 0.50 & $10^2$ & $-141.8$ \\ \hline
\end{tabular}
\label{tab:GA_runs}
\end{table}
It can be seen from this that the best individual, or rather the kernel, came from trial 4 with the largest mutation rate of 50\% (trial 4). However, credit must also be given to the number of generations which is also necessary to complement the mutation rate in producing a population with stronger, fitter, individuals. The fittest individuals per trial in Table \ref{tab:GA_runs} are characterized in Table \ref{tab:GA_kernels}.

\begin{table}[h!]
\center
\caption{Characterization of the fittest individuals emerging from the four GP-GA trial runs presented in Table \ref{tab:GA_runs} together with their corresponding $H_0$ prediction, log-marginal likelihood, $\chi^2$, fitness, and penalty.}
\begin{tabular}{| c | c | c | c | c | c |}
\hline
$\phantom{ \dfrac{1}{1} }$ kernel $\phantom{ \dfrac{1}{1} }$ & $H_0$ $\left[\text{km} \ \text{s}^{-1} \text{Mpc}^{-1} \right]$ & $\ln \mathcal{L}$ & $\chi^2$ & fitness & penalty \\ \hline
$\phantom{\dfrac{1}{1}}$ \textit{Hybrid} RBF-RQ $\phantom{\dfrac{1}{1}}$ & $70.6 \pm 5.5$ & $-131.49$ & 13.1 & $-143.5$ & 12.0  \\ \hline
$\phantom{\dfrac{1}{1}}$ \textit{Hybrid} RBF-RQ-M52 $\phantom{\dfrac{1}{1}}$ & $66.9 \pm 6.3$ & $-131.38$ & 12.0 & $-148.5$ & 17.2 \\ \hline
$\phantom{\dfrac{1}{1}}$ \textit{Mostly} RQ $\phantom{\dfrac{1}{1}}$ & $66.7 \pm 6.4$ & $-131.36$ & 11.7 & $-143.4$ & 12.0 \\ \hline
$\phantom{\dfrac{1}{1}}$ \textit{Hybrid} RBF-M52 $\phantom{\dfrac{1}{1}}$ & $69.8 \pm 5.8$ & $-131.48$ & 12.7 & $-141.8$ & 10.3 \\ \hline
\end{tabular}
\label{tab:GA_kernels}
\end{table}

It is shown here that the fittest individuals always come from at least a mixture of two kernels, the most special one having a fitness of $\mathcal{F} = -141.8$, characterized by the RBF and M52 subkernels. The hyperparameters of this \textit{Hybrid} RBF-M52 kernel are given by
\begin{equation}
\left( C_{\text{RBF}}, l_{\text{RBF}} , n_{\text{RBF}}, C_{\text{M52}}, l_{\text{M52}}, n_{\text{M52}} \right) = \left( 13.81, 1.37, 4.90, 126.10, 2.14, 0.44 \right)\,.
\end{equation}
This surpassed the fitness of the \textit{Mostly} RQ kernel because it suffered from less penalty. The Mostly RQ is a mixture of the RBF and RQ kernels $(k_\text{eff} = 7)$ but with the RBF contributions being subdominant compared to those coming from the RQ side. The above results show that the GP-GA implementation have a preference towards simplicity or a smaller number of hyperparameters.

\begin{figure}[h!]
\center
\includegraphics[width = 0.5 \textwidth]{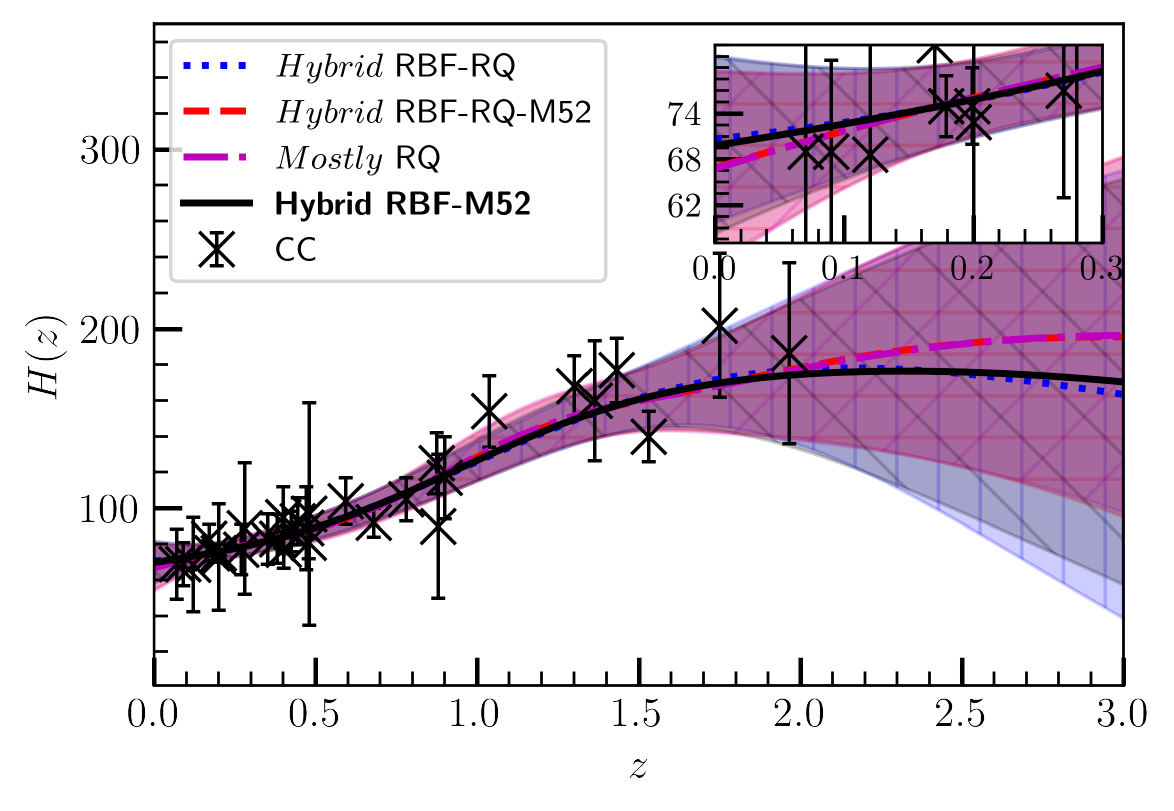}
\caption{The GP-GA reconstructions of the Hubble function from the cosmic chronometers data set for the naturally-selected kernels presented in Table \ref{tab:GA_kernels}. The colored-hatched regions show the $2\sigma$ region around the mean. Hatches used: (Hybrid RBF-RQ, $'|'$), (Hybrid RBF-RQ-M52, $'-'$), (Mostly RQ, $'/'$), (Hybrid RBF-M52, $' \text{\textbackslash} '$).}
\label{fig:Hz_gp_ga}
\end{figure}

The reconstructed Hubble functions based on the kernels presented in Table \ref{tab:GA_kernels} are shown in Fig.~\ref{fig:Hz_gp_ga}. It can be seen here that there is a notable overlap between the Hybrid RBF-RQ-M52 and the Mostly RQ predictions. The same can be said of the predictions of the Hybrid RBF-RQ and the fittest Hybrid-RBF-M52. Looking back at Table \ref{tab:GA_kernels}, this agreement can be traced back to the log-marginal likelihood of the corresponding kernels. In this context, the fitter individuals do not come out as having the largest log-marginal likelihood but rather as ones striking a balance between simplicity and log-marginal likelihood.

\begin{table}[h!]
\center
\caption{Parameters and best fitness for four trials with the GA with $n_\text{gen}$ generations for the SNe data set reconstruction.}
\begin{tabular}{| c | c | c | c | c | c |}
\hline
trial & population size & selection rate & mutation rate & $n_\text{gen}$ & best fitness \\ \hline
$\phantom{\dfrac{1}{1}}$ 1 $\phantom{\dfrac{1}{1}}$ & $10^4$ & 0.5 & 0.15 & $10^1$ & $44.0$  \\ \hline
$\phantom{\dfrac{1}{1}}$ 2 $\phantom{\dfrac{1}{1}}$ & $10^4$ & 0.3 & 0.30 & $10^1$ & $49.3$ \\ \hline
$\phantom{\dfrac{1}{1}}$ 3 $\phantom{\dfrac{1}{1}}$ & $10^3$ & 0.1 & 0.10 & $10^2$ & $45.0$ \\ \hline
$\phantom{\dfrac{1}{1}}$ 4 $\phantom{\dfrac{1}{1}}$ & $10^3$ & 0.3 & 0.50 & $10^2$ & $51.3$ \\ \hline
\end{tabular}
\label{tab:GA_runs_sne}
\end{table}

Now, Tables \ref{tab:GA_runs_sne} and \ref{tab:GA_kernels_sne} show the analogous trial runs for the SNe data set. Clearly, Table \ref{tab:GA_runs_sne} echoes the earlier sentiment on the importance of mutation in producing a more competitive, fitter offspring in the population. In this case, the GA with the strongest mutation (trial 4) was able to produce the fittest individual in notably less than half the run time of the ones with larger populations. The fittest individuals per trial in Table \ref{tab:GA_runs_sne} are characterized in Table \ref{tab:GA_kernels_sne}. 

\begin{table}[h!]
\center
\caption{Characterization of the fittest individuals emerging from the four GP-GA trial runs presented in Table \ref{tab:GA_runs_sne} together with their log-marginal likelihood, $\chi^2$, fitness, and penalty.}
\begin{tabular}{| c | c | c | c | c | c |}
\hline
$\phantom{ \dfrac{1}{1} }$ kernel $\phantom{ \dfrac{1}{1} }$ & $\ln \mathcal{L}$ & $\chi^2$ & fitness & penalty \\ \hline
$\phantom{\dfrac{1}{1}}$ \textit{Hybrid} RBF-RQ-M52 $\phantom{\dfrac{1}{1}}$ & $62.4$ & 33.4 & $44.0$ & 18.4  \\ \hline
$\phantom{\dfrac{1}{1}}$ \textit{Mostly} M52 $\phantom{\dfrac{1}{1}}$ & $62.3$ & 32.3 & $49.3$ & 12.9 \\ \hline
$\phantom{\dfrac{1}{1}}$ \textit{Mostly} RBF-M52 $\phantom{\dfrac{1}{1}}$ & $63.4$ & 31.7 & $45.0$ & 18.4 \\ \hline
$\phantom{\dfrac{1}{1}}$ \textit{Hybrid} RBF-M52 $\phantom{\dfrac{1}{1}}$ & $62.4$ & 23.8 & $51.3$ & 11.1 \\ \hline
\end{tabular}
\label{tab:GA_kernels_sne}
\end{table}

Table \ref{tab:GA_kernels_sne} also shows that even for the SNe data the fittest individuals come as a mixture of at least two of the basic subkernels. The most special one, \textit{Hybrid} RBF-M52, a mixture of the RBF and M52 kernels, has a fitness of $\mathcal{F} = 51.3$ and is uniquely described by the chromosome
\begin{equation}
\left( C_{\text{RBF}}, l_{\text{RBF}} , n_{\text{RBF}}, C_{\text{M52}}, l_{\text{M52}}, n_{\text{M52}} \right) = \left( 47.68, 57.17, 4.17, 12.81, 35.81, 2.92 \right)\,.
\end{equation}
The \textit{Hybrid} RBF-M52 bests the \textit{Mostly} M52 kernel, a mixture of M52 and a subdominant RQ, because of the penalty function. Understandably, any kernel with an effective RQ part will always have an extra four hyperparameters in $k_\text{eff}$ and so pay the price for complexity. On the other hand, both the \textit{Hybrid} RBF-RQ-M52 and the \text{Mostly} RBF-M52 are mixtures of the RBF, RQ, and M52 kernels; the latter instead having subdominant RQ contributions compared to the other two. Therefore, we find that our GP-GA implementation supports a balance between simplicity and the effectiveness of the reconstruction.

We want to also point out two important observations at this point. Firstly, consider the intriguing result of obtaining the same fittest kernel for two independent data sets. In this case, it turned out that a mixture of the RBF and M52 kernels seem to best represent \textit{both} the CC and SNe data sets. In general, there is no \textit{a priori} reason for a single type kernel to be able to describe different observations. This may of course be considered to be merely an outstanding coincidence. This may also be the result of the RBF kernel being the most natural choice of kernel for GP and the combination with M52 producing the best balance of kernel hyperparameters against the fitness function (Eq. (\ref{eq:fitness_function})). Second, Table \ref{tab:GA_kernels_sne} shows a prime example of when a larger log-marginal likelihood does not anymore correspond to a smaller $\chi^2$. Care must then be given when making a connection between the two statistics.

\begin{figure}[h!]
\center
\includegraphics[width = 0.5 \textwidth]{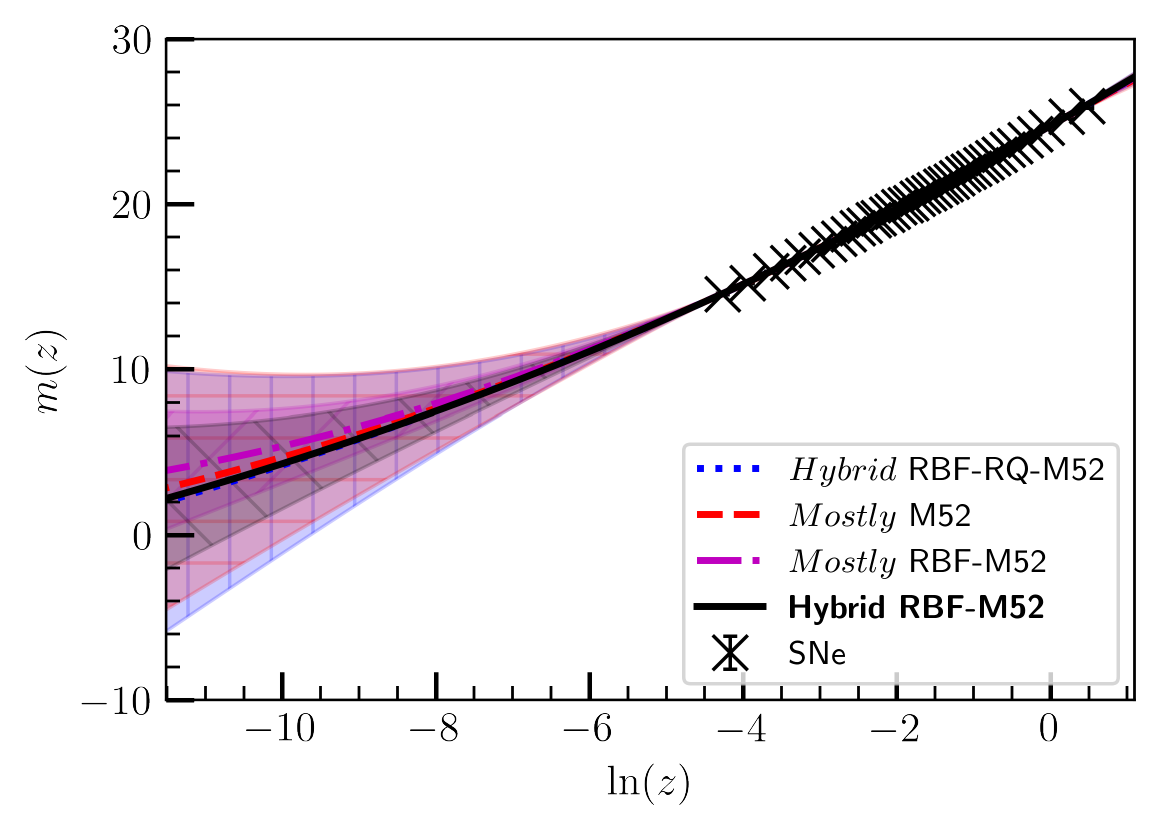}
\caption{The GP-GA reconstructions of the SNe apparent magnitudes from the Pantheon data set for the naturally-selected kernels presented in Table \ref{tab:GA_kernels_sne}. The colored-hatched regions show the $2\sigma$ region around the mean. Hatches used: (Hybrid RBF-RQ, $'|'$), (Hybrid RBF-RQ-M52, $'-'$), (Mostly RQ, $'/'$), (Hybrid RBF-M52, $' \text{\textbackslash} '$).}
\label{fig:mz_gp_ga}
\end{figure}

The GP reconstructions of the SNe apparent magnitudes for the trials characterized in Tables \ref{tab:GA_runs_sne} and \ref{tab:GA_kernels_sne} are shown in Fig. \ref{fig:mz_gp_ga}. Similar to earlier results, we find that any of the reconstructions can almost be visually-accepted. In fact, in this case of the SNe reconstruction, the GPs are practically indistinguishable in the places where the data points are clustered, i.e., $\ln(z) \gtrsim -5$. The combined GP and GA implementation of this section instead showcases a way to find the one kernel which best balances simplicity and representation of the data.

To end, we emphasize that GA is a stochastic process, i.e., trials with the same parameters generally lead to different results. See more information in the Appendix \ref{sec:ga_closer_look}. This means that it will not be surprising to find kernels with better fitness than the ones singled out in this section provided more computational time and resources. The stochastic nature may potentially be integrated with Bayesian tools such as the ABC-SMC. The resulting method will surely be something to look forward to.

\section{Conclusion}
\label{sec:conclusions}

In this paper, we have approached the GP kernel selection problem with (1) approximate Bayesian computation with sequential Monte Carlo sampling, and (2) genetic algorithm as a means to address this often arbitrary choice of the kernel (and used the overall resulting method to reconstruct the cosmic chronometers and the Pantheon SNe data sets). The GP-ABC-SMC reconstruction pointed to the Mat\'{e}rn$\left( \nu = 5/2 \right)$ kernel as being preferable among the two-hyperparameter family of kernels. On the other hand, the GP-GA reconstruction singled out a hybrid of the RBF and M52 kernels as the fittest, threading the best balance between simplicity and a preference of a larger log-marginal likelihood.

We emphasize that the results obtained using the methods of this paper may be dependent on the data sets under consideration. Taking from this view, it is most interesting that the Matern$\left( \nu = 5/2 \right)$ kernel was singled out for both the CC and SNe data sets in the GP-ABC-SMC reconstruction. Also, the same hybrid of the RBF and M52 kernels notably emerged as the fittest in the GP-GA reconstruction of both data sets. This may be a coincidence, or simply a result of both data sets sharing an underlying indication of an expanding Universe. Nonetheless, it will interesting to see if the above kernels can continue to be competitive in this content among the population of kernels when different cosmological observations are used, e.g., baryon acoustic oscillations and $f \sigma_8$ data.

Another important point to raise is that the ultimate goal within the kernel selection problem is to remove the subjectivity in the choice of kernel for a GP given a particular data set. This was accomplished in the integration of the GP with the ABC-SMC and the GA. Both methods automate the kernel selection and so overcomes the kernel prejudices entrusted to the user in the vanilla implementation of the GP. However, we have still chosen a fitness function strategy. We have attempted to motivate the naturalness of this choice but there may be future analyses that produce a closer approximation to an implementation independent of this choice.

As future work, further improvements to the general approach to GA regression can still be advanced. A combination of GP, ABC-SMC, and the GA may potentially be implemented. This may be a good way to extend the ABC-type analysis to more complex kernels without sacrificing too much computational budget. Also, it will be interesting to further explore the more intimate connection between the log-marginal likelihood and the $\chi^2$. The results presented here suggest that in most times, increasing the log-marginal likelihood decreases the $\chi^2$ fit. However, the opposite relation is mostly not true and it will be useful to know when one might draw the line between the two measures.

Lastly, one of the important motivations for the development of Hubble data reconstruction methods is to study the fundamental physics behind dark energy. A most natural application of the results of this paper is then to constraining the dark energy equation of state or the potentials describing an alternative theory of gravity. Such practical considerations already exist throughout the recent cosmology literature, and may now be revisited without the need to prefer any single unnatural choice of the covariance function.

\begin{acknowledgments}\label{sec:acknowledgements}
The authors would like to thank Reina Reyes, Johnrob Bantang, and Ahmed Gad for helpful conversations.
JLS would like to acknowledge networking support by the COST Action CA18108 and funding support from Cosmology@MALTA which is supported by the University of Malta. JLS would also like to acknowledge funding from ``The Malta Council for Science and Technology'' in project IPAS-2020-007.
\end{acknowledgments}

\appendix

\section{Genetic algorithm: A closer look}
\label{sec:ga_closer_look}

\begin{figure}[h!]
\center
	\subfigure[ ]{
		\includegraphics[width = 0.47 \textwidth]{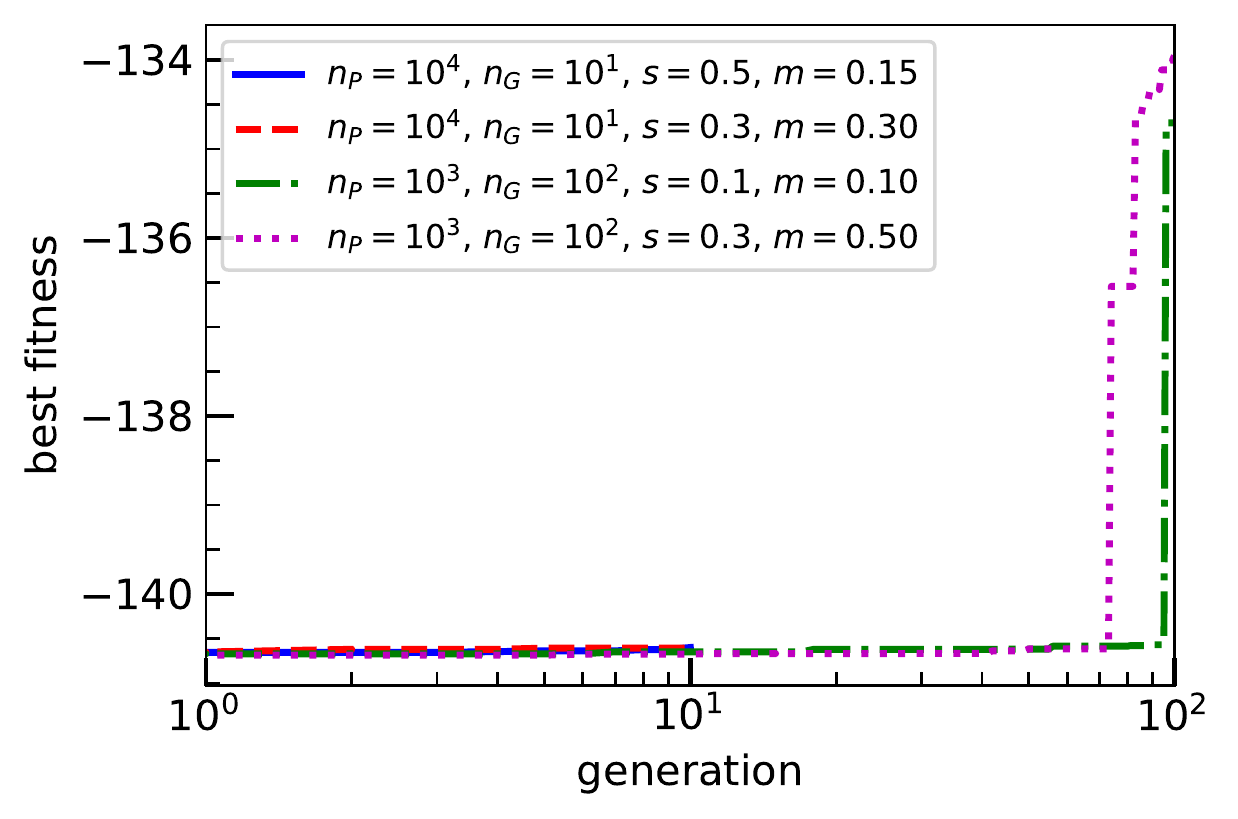}
		}
	\subfigure[ ]{
		\includegraphics[width = 0.47 \textwidth]{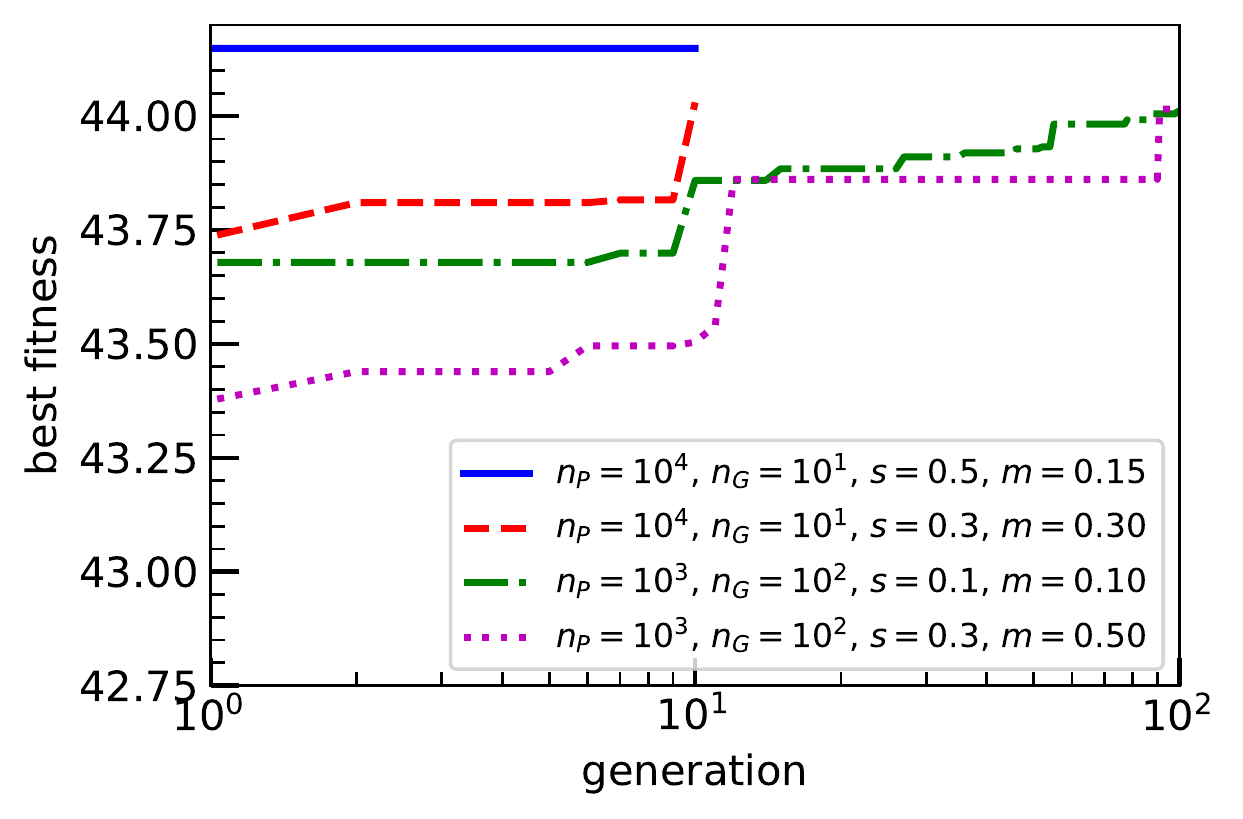}
		}
\caption{Best fitness per generation for trials with different GA parameters for the (a) CC and (b) SNe data sets. The $n_P$, $n_G$, $s$, and $m$ are the population size, number of generations, selection rate, and mutation probability, respectively.}
\label{fig:bfit_gen}
\end{figure}

In this section, we take a closer look at what happens during the GA for different parameters. Each trial takes about an hour to two to complete in a standard 8 GB RAM laptop with an Intel Core I7 processor. The resulting evolution of the best fitness is shown in Fig. \ref{fig:bfit_gen}.

The two plots visually-describe the interplay between the different parameters entering the GA. Clearly, in both cases, it can be be seen that mutation plays the largest role in the optimization process. Without it, the algorithm is only more likely to get stuck in a local maxima of the fitness function. This is undesirable for problems with multimodal fitness functions, such is the particular case in Sec. \ref{sec:gp_ga}, and makes the case of mutation as an irreplaceable feature of the GA in such applications.
Fig. \ref{fig:bfit_gen} supports this. In both the CC and SNe applications, the best fitness per generation can be seen to evolve more efficiently with higher mutation rates regardless of the population size or the initial population. The parameters in Figure \ref{fig:bfit_gen} may also be observed to be the same ones in Tables \ref{tab:GA_runs} and \ref{tab:GA_runs_sne}. However, the best fitness values are different. This is understandably reflective of the stochastic nature of the GA.

Other factors such as the selection, crossover, and mutation mechanisms can also affect the efficiency of the GA optimization. A detailed discussion of these is beyond the scope of this paper. We draw the interested reader to the python package \textit{pygad} \cite{pygad}.


\providecommand{\href}[2]{#2}\begingroup\raggedright\endgroup

\end{document}